%% file: main.tex
\documentclass{sig-alternate}

\usepackage{url}
\usepackage{graphicx}
	\graphicspath{{./images/}}
	\DeclareGraphicsExtensions{.png,.pdf}
\usepackage{amsmath}
\usepackage{paralist}
\usepackage{color}
\usepackage{xspace}

\newcommand{\sedano}{\textsc{Sedano}\xspace}
\newcommand{\fakeparagraph}[1]{\noindent\textbf{#1}}

  \newif\ifdraft
  \drafttrue

\begin{document}


\CopyrightYear{2016}
\setcopyright{rightsretained}
\conferenceinfo{SIGIR '16}{July 17-21, 2016, Pisa, Italy} 
\isbn{978-1-4503-4069-4/16/07} 
\doi{http://dx.doi.org/10.1145/2911451.2926730}

\clubpenalty=10000 
\widowpenalty = 10000



%

\title{Sedano: A News Stream Processor for Business}
%
%
%
%
%

\numberofauthors{1} 
%
\author{
\alignauthor Ugo Scaiella, Giacomo Berardi, Giuliano Mega, Roberto Santoro\\
       \affaddr{Spaziodati, Italy}\\
       \email{\{scaiella,berardi,mega,santoro\}@spaziodati.eu}
}

\maketitle
\begin{abstract}
We present \sedano, a system for processing and indexing a continuous stream of business-related news. \sedano defines pipelines whose stages analyze and enrich news items (e.g., newspaper articles and press releases). News data coming from several content sources are stored, processed and then indexed in order to be consumed by Atoka, our business intelligence product. Atoka users can retrieve news about specific companies, filtering according to various facets.
\sedano features both an entity-linking phase, which finds mentions of companies in news, and a classification phase, which classifies news according to a set of business events. Its flexible architecture allows \sedano to be deployed on commodity machines while being scalable and fault-tolerant.
\end{abstract}

%
%


%
%

%
%
\printccsdesc



\input{content}
\end{document}

%% file: content.tex
\section{Introduction}
Nowadays a deluge of news articles and press releases often bewilders readers; this explains why news aggregators have surged in popularity. While news aggregators do a good job for readers in general, they may not be enough if someone is interested in specific business-related news and events, e.g.: which are the news where a particular company is mentioned? Which news recently talked about company events, like lay-offs or mergers and acquisitions? Which are the interesting events regarding a specific company?
For this reason we created \sedano, a system that is able to ingest, process and enrich news stream in a scalable manner. The news are indexed and offered via a RESTful API, in order to be consumed by our end-user product Atoka\footnote{\url{http://atoka.io}}, a semantic tool for \emph{lead generation}, but also by other business clients that integrate the information provided by \sedano in their products. 


\section{Architecture}
\begin{figure}
	\centering
    \includegraphics[scale=0.5]{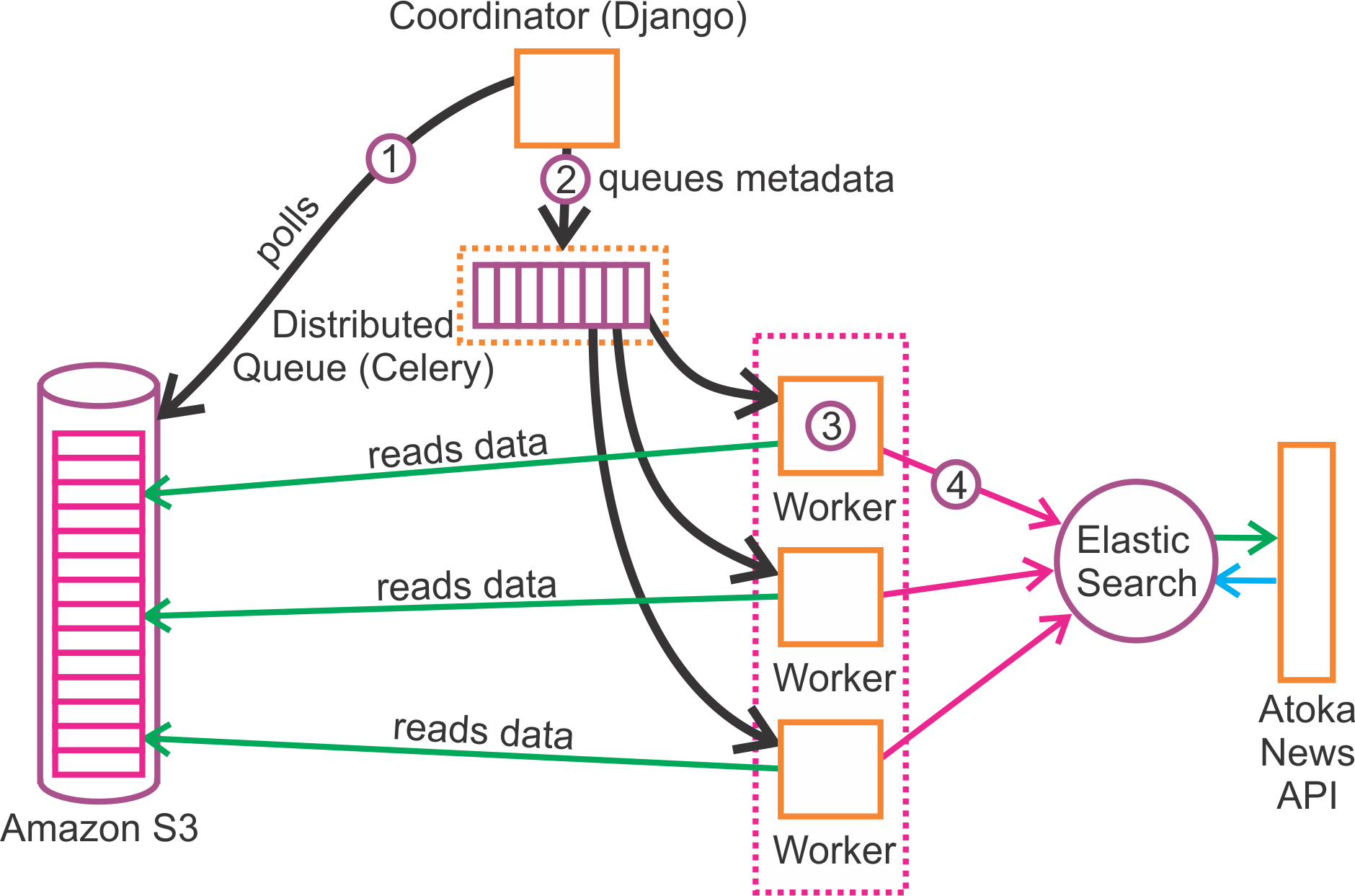}
    \caption{\sedano's base architecture.}
    \label{fig:architecture}
\end{figure}

\sedano is based on a simple distributed architecture, depicted in Figure~\ref{fig:architecture}. It leverages Amazon's S3\footnote{\url{http://aws.amazon.com/s3}} for high aggregated throughput, and the Celery\footnote{\url{http://www.celeryproject.org}} distributed queue for control messages. Data is stored in S3 as a collection of \emph{data chunks}, i.e., S3 objects containing \emph{news items}. Processing in \sedano is triggered by a \emph{coordinator node}, which \begin{inparaenum}[(1)]
	\item periodically (i.e., every minute) polls S3 for new chunks and,
    \item enqueues \emph{work orders} referring to chunks that require processing into Celery. Once into Celery, work orders become visible to \emph{worker nodes} who then dequeue them, 
    \item download the corresponding chunks in parallel from S3, and process them independently. At the end of the processing, the resulting news items are 
    \item indexed into an ElasticSearch\footnote{\url{http://www.elastic.co}} cluster.
\end{inparaenum}
Basic fault tolerance is achieved by having Amazon's Elastic Container Service (ECS)\footnote{\url{https://aws.amazon.com/ecs}} supervise individual components, respawning them into available machines in the event of crashes.

Processing at worker nodes takes place as a series of sequential stages that make up \sedano's \emph{processing pipelines}. Pipelines are composed of three major steps, discussed next.

\fakeparagraph{Cleansing and normalization.} The first step at every pipeline deals with content cleansing and normalization---we extract the relevant content from each news item by processing the HTML in which they are represented, and compute several measures (e.g. the length of an article or if its content is well-structured) which later help us understand if that news item is ``good'' or not (and therefore whether or not it should be served to certain clients).

\fakeparagraph{Deduplication.} News items are then deduplicated. Since finding duplicates by comparing new documents with previously indexed ones would be costly, we adopt a more scalable on-line approach, based on a novel Locality-Sensitive Hashing (LSH) algorithm which mixes the Solr\footnote{\url{http://lucene.apache.org/solr}} approach with the Nilsimsa algorithm\footnote{\url{https://en.wikipedia.org/wiki/Nilsimsa_Hash}} to hash similar news items into the same code. A news item is then tagged as a duplicate if and only if another news item with the same hash code is known to have been processed before. To keep this information consistent across worker nodes, we rely on a scalable key-value store with strong consistency guarantees.

\fakeparagraph{Enrichment.} The remainder of this section discusses the two main enrichment components in the \sedano pipelines: the \emph{Dandelion Company API} (Section~\ref{sec:dandelion}), an entity linking platform focused on the identification of company mentions in news, and \emph{Selino} (Section~\ref{sec:selino}), an ensemble of linear multi-label text classifiers whose classes are business events of interest.


\subsection{Dandelion Company API}\label{sec:dandelion}

Dandelion API\footnote{\url{https://dandelion.eu}} is a platform for text-analytics as a service: it is the evolution of {\sc Tagme}~\cite{tagme}, a state-of-the-art entity linking system based on a knowledge graph extracted from Wikipedia, whose main benefits are its performance on short texts and its speed. Our partnership with Cerved\footnote{\url{https://www.cerved.com}}, the leader of business information in Italy, gave us access to data of all Italian companies. We used this data to specialize the entity linking system into the Dandelion Company API for identifying company mentions. It includes information on almost 3M companies, which are added as new entities to our knowledge graph. Despite the huge number of entities that have been added, the system preserves its speed.

Since the new entities do not have cleaned data like Wiki\-pedia ones, they may introduce noise. For this reason we added a new layer, based on a Named Entity Recognition (NER) classifier, which is able to identify named entities of type ``company''. The result of the NER classifier is used as an additional feature for the subsequent disambiguation step\footnote{Refer to \cite{tagme} for further details and terminology}, in order to better identify company entities.

We carried out several tests with data sets created by means of crowd-sourced annotations, and we yield a $F_{1}$ of 60\%. However, for each annotation we also compute a confidence score; setting a threshold for this score can be used to balance prevision vs. recall. Since for this kind of product we are mainly interested in precision, we decided to set the threshold to maximize the $F_{0.5}$, yielding a precision of $\sim 80\%$ and a recall of $\sim 45\%$.


\subsection{Selino}\label{sec:selino}

Selino is a component of the \sedano pipeline dedicated to the automatic classification of news according to a predefined set of categories. The categories we are interested in are business events with negative connotation (i.e.: layoffs, strikes, shutdowns, material damages, financial losses, frauds and legal issues) plus other generic events (mergers and acquisitions, new product launches and management changes).

We first built a training set for each category. Given the small size of the taxonomy we created an independent data set for each category; we chose CrowdFlower\footnote{\url{http://www.crowdflower.com}} as crowdsourcing platform for the manual annotation of news.
Instead of randomly sampling news for each annotation job, we selected sets of articles ($\sim12K$ samples per category) through a boolean retrieval model. Each category query was made by expanding a seed query of event related terms. The expansion was performed by (a) creating a vector space model of Italian word semantics using Word2Vec~\cite{word2vec}, (b) identifying the most similar words to the seeds, (c) manually selecting words for defining the query.
This approach allowed us to minimize the manual annotation work and to obtain balanced training sets (each one is made of $\sim30\%$ positive samples).

Our approach to event detection is through supervised learning: we chose a logistic regression algorithm as text classifier. This linear model has proven to be effective for text classification~\cite{textLogisticRegression} and it is also reasonably fast.
We evaluated the classifiers through 3-fold cross validation, obtaining good results (i.e. an average precision of $0.8$ and an average recall of $0.6$), especially on precision. The latter is our critical evaluation measure, since a false positive on a negative event can produce unpleasant image damages to the companies cited in an article.


\section{Conclusions}

\sedano is an effective system for processing news and extracting rich information about companies and interesting signals for the business domain. We have shown its design and components, among which the main ones are the entity linking system for identifying company mentions and the events detection classifier. \sedano has proven to be reliable and scalable, analyzing and indexing hundreds of thousands of Italian news items every day.

We plan to improve further this application by:

\begin{compactitem}
\item Integrating new components, such as a clustering algorithm to group together related news.
\item Moving from the current static event-detection models to an on-line active learning approach, in which we sample news with ``ambiguous'' classifications and we annotate them through CrowdFlower.
\item Extracting specific facts about events from textual contents, employing more sophisticated event detection techniques.
\end{compactitem}